# Controlling Entanglement Dynamics by Choosing Appropriate Ratio between Cavity-Fiber Coupling and Atom-Cavity Coupling

Liao Chang-Geng,† Yang Zhen-Biao, Chen Zi-Hong, and Luo Cheng-Li

Department of Electronic Science and Applied Physics, Fuzhou University Fuzhou 350002, China

Abstract

The entanglement characteristics including the so-called sudden death effect between two identical two-level atoms trapped in two separate cavities connected by an optical fiber are studied. The results show that the time evolution of entanglement is sensitive not only to the degree of entanglement of the initial state but also to the ratio between cavity-fiber coupling ($v$) and atom-cavity coupling ($g$). This means that the entanglement dynamics can be controlled by choosing specific $v$ and $g$.

## Ⅰ Introduction

Entanglement, one of the most striking features of quantum mechanics, plays a key role in quantum computation and quantum information processing (QIP) [1~2]. In recent years, there has been an ongoing effort to characterize qualitatively and quantitatively the entanglement properties and apply then in the field. However, in the process of entanglement distribution and manipulation, each qubit is unavoidably exposed to its environment. This leads to local decoherence which will sooner or later spoil the entanglement. The fragility of nonlocal quantum coherence is recognized as a main obstacle to realizing QIP [3, 4]. What's worse, even under the influence of pure vacuum noise two entangled qubits may become completely disentangled in a finite time, an interesting phenomenon termed entanglement sudden death (ESD) [5]. It has been found in numerous theoretical examinations to occur in a wide variety of entanglements involving pairs of atomic, photonic, and spin qubits, continuous Gaussian states, and subsets of multiple qubits and spin chains [6]. Recently, it is reported that the ESD phenomenon has been observed in a quantum optics experiment [7]. Therefore, special circumstances are needed to see "anti-ESD," the creation or rebirth of entanglement from disentangled states. It has been pointed out that the cavity coherent state can be used to control the ESD and entanglement sudden birth （ESB） in cavity quantum electrodynamics (QED) [8]. In Ref. [9], Yamamoto et al. have shown that the direct measurement feedback method can avoid ESD, and further, enhance the entanglement. More recently, it is shown that the time of ESD can be controlled by the classical driving fields [10]. On the other hand, several theoretical schemes based on fiber-connected cavity modes for realizing distributed QIP have been proposed [11~24], it would be very interesting to see the entanglement dynamics of this system which may be different from previous ones.

In the present paper, the entanglement dynamics of a quantum system (as is shown in Fig.1) consisting of two identical two-level atoms trapped in two separate cavities connected by an optical fiber is studied by means of Wootters's concurrence[25]. The results show that the time evolution of entanglement is sensitive not only to the degree of entanglement of the initial state but also to the ratio between cavity-fiber coupling ($v$) and atom-cavity coupling ($g$). This means that the entanglement evolvement can be controlled by specific $v$ and $g$.

The paper is organized as follows. In Sec.Ⅱ, we present the model and its theoretical description. In Sec.Ⅲ, we show how entanglement characteristics evolve. The influence of the ratio between cavity-fiber coupling ($v$) and atom-cavity coupling ($g$) and the degree of entanglement of the initial state on the ESD is also studied. In Sec.Ⅳ, we give the conclusion.

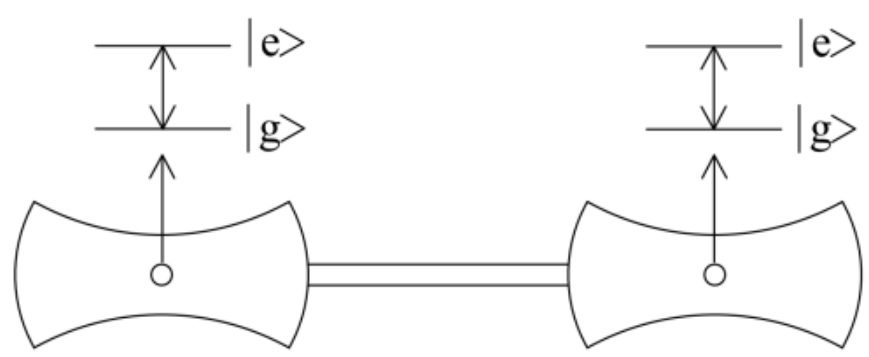

Fig.1 Experimental setup.

## Ⅱ Model

As shown in Fig.1, two completely identical two-level atoms are trapped in two distant high-finesse optical cavities, which are connected by an optical fiber. The atoms have an excited state $|e\rangle$ and a ground state $|g\rangle$. The transition $|e\rangle \leftrightarrow |g\rangle$ of each atom is resonant with the local cavity mode. The interaction Hamiltonian in rotating wave approximation can be written as $H_i = g_i a_i S_i^+ + \mathrm{H.c.}$, where $g_i$ is the coupling strength between the mode of cavity $i$ and the trapped atom. For simplicity, we assume that each $g_i$ has the same value, thus we will use $g$ instead of it in the following paper, $a_i$ is the annihilation operator for photons in the mode of cavity $i$, and $S_i^+ = |e_i\rangle\langle g_i|$ is the atomic raising operator for the atom in cavity $i$.

The coupling between the cavity and fiber modes is modeled by the interaction Hamiltonian $H_{if} = \sum_{k=1}^{\infty} v_k \{b_k[a_1^+ + (-1)^k a_2^+] + \mathrm{H.c.}\}$ [12], where $b_k$ is the annihilation operator for the fiber mode $k$ and $v_k$ is the coupling strength of the cavity mode to the fiber mode $k$. The factor $(-1)^k$ denotes the phase difference between the adjacent modes at the fiber end. In the short fiber limit $2l\bar{v}/(2\pi C) \leq 1$ [12], where $l$ is the length of fiber and $\bar{v}$ is the decay rate of the cavity fields into a continuum of fiber modes, only one resonant mode $b$ of the fiber interacts with the cavity modes. Therefore, for this case, the Hamiltonian $H_{if}$ reduces to $H_{if} = v[b(a_1^+ + a_2^+) + \mathrm{H.c.}]$.

In the interaction picture, the total Hamiltonian of the atom-cavity-fiber combined system is [12, 13]

$$H_i=\sum_{i=1}^{2}g(a_iS_i^{+}+\text{H.c.})+v[b(a_1^{+}+a_2^{+})+\text{H.c.}]. \tag{1}$$

Suppose that at the initial time all modes of both the cavity and fiber fields are in the vacuum state $|000\rangle$, with $|000\rangle\equiv|0\rangle_{c1}|0\rangle_{f}|0\rangle_{c2}$, where the subscript $c1$, $c2$ and $f$ denote two cavities and the connected fiber. The atoms are assumed to be in the state

$$|\phi_{atoms}\rangle=\cos\theta|e\rangle_1|g\rangle_2+\sin\theta|g\rangle_1|e\rangle_2, \tag{2}$$

where the subscript 1 and 2 denote the corresponding atom in cavity $c1$ and $c2$. Hence, the total state of the system at $t=0$ is

$$|\psi(0)\rangle=\cos\theta|eg000\rangle+\sin\theta|ge000\rangle. \tag{3}$$

The time evolution of the system in this initial state can be expressed as

$$|\psi(t)\rangle=N_1|eg000\rangle+N_2|gg100\rangle+N_3|gg010\rangle+N_4|gg001\rangle+N_5|ge000\rangle, \tag{4}$$

where

$$N_1=\frac{2v^2+g^2\cos\sqrt{2v^2+g^2}t}{4v^2+2g^2}(\cos\theta+\sin\theta)+\frac{\cos gt}{2}(\cos\theta-\sin\theta), \tag{5}$$

$$N_2=-\frac{ig\sin\sqrt{2v^2+g^2}t}{2\sqrt{2v^2+g^2}}(\cos\theta+\sin\theta)-\frac{i\sin gt}{2}(\cos\theta-\sin\theta), \tag{6}$$

$$N_3=-\frac{gv}{2v^2+g^2}(1-\cos\sqrt{2v^2+g^2}t)(\cos\theta+\sin\theta), \tag{7}$$

$$N_4 = -\frac{ig\sin\sqrt{2v^2+g^2}t}{2\sqrt{2v^2+g^2}}(\cos\theta+\sin\theta)+\frac{i\sin gt}{2}(\cos\theta-\sin\theta)\,, \tag{8}$$

$$N_5 = \frac{2v^2+g^2\cos\sqrt{2v^2+g^2}t}{4v^2+2g^2}(\cos\theta+\sin\theta)-\frac{\cos gt}{2}(\cos\theta-\sin\theta)\,. \tag{9}$$

Let $\rho(t)=|\psi(t)\rangle\langle\psi(t)|$ be a density matrix of the above state, the reduced density matrix $\rho^{atoms}(t)$ for the atoms at any time $t$ can be obtained by tracing out over the degrees of freedom of the cavity and fiber fields. In the standard basis $\{|ee\rangle,|eg\rangle,|ge\rangle,|gg\rangle\}$, it can be easily expressed as

$$\rho^{atoms}(t)=\begin{pmatrix} 0 & 0 & 0 & 0 \\ 0 & \rho_{22}(t) & \rho_{23}(t) & 0 \\ 0 & \rho_{32}(t) & \rho_{33}(t) & 0 \\ 0 & 0 & 0 & \rho_{44}(t) \end{pmatrix}, \tag{10}$$

with

$$\rho_{22}(t)=|N_1|^2\,,\rho_{23}(t)=N_1N_5^*\,,\rho_{32}(t)=N_5N_1^*\,,\rho_{33}(t)=|N_5|^2\,,\rho_{44}(t)=|N_2|^2+|N_3|^2+|N_4|^2\,.$$

## Ⅲ Entanglement measure and entanglement dynamics

In two-qubit domains, there are several good measures to quantify the degree of entanglement. Throughout the paper we will use Wootters's concurrence [25] $C(\rho)$ which is conveniently defined for both pure and mixed states.

The concurrence $C(\rho)$ for the reduced density matrix $\rho$ of a two-qubit system is defined as $C(\rho)=\max\{0,\sqrt{\lambda_1}-\sqrt{\lambda_2}-\sqrt{\lambda_3}-\sqrt{\lambda_4}\}$, where the quantities $\lambda_i$ are the eigenvalues of the matrix $\zeta$ :

$$\zeta = \rho(\sigma_y \otimes \sigma_y)\rho^*(\sigma_y \otimes \sigma_y), \qquad (11)$$

arranged in decreasing order. Here $\sigma_y$ is the usual Pauli matrix and $\rho^*$ denotes the complex conjugation of $\rho$. It can be shown that the concurrence varies from $C=0$ for a disentangled state to $C=1$ for a maximally entangled state.

From (5), (9), (10) and (11), the concurrence for the reduced density matrix $\rho^{atoms}(t)$ can be easily derived as

$$C(\rho^{atoms}(t)) = 2\left|[\frac{2v^2+g^2\cos\sqrt{2v^2+g^2}t}{4v^2+2g^2}(\cos\theta+\sin\theta)]^2-[\frac{\cos gt}{2}(\cos\theta-\sin\theta)]^2\right|. \qquad (12)$$

Suppose that $r=\frac{v}{g}$, the concurrence (12) reduces to

$$C(\rho^{atoms}(t)) = 2\left|[\frac{2r^2+\cos\sqrt{2r^2+1}gt}{4r^2+2}(\cos\theta+\sin\theta)]^2-[\frac{\cos gt}{2}(\cos\theta-\sin\theta)]^2\right|. \qquad (13)$$

(1) For $\theta=-\frac{\pi}{4}$, which the atoms are initially prepared in the maximal entanglement state

$$|\psi(0)\rangle = \frac{\sqrt{2}}{2}(|eg\rangle-|ge\rangle), \qquad (14)$$

we can get the concurrence for atom-atom entanglement at time $t$ with the above case.

$$C(\rho^{atoms}(t)) = \cos^2 gt. \qquad (15)$$

From Eq. (15) we can easily find that the pairwise entanglements of two atoms exhibit ESD and the time evolution of atom-atom entanglement is irrelevant to the cavity-fiber coupling ($v$). The entanglements of atoms evolve periodically in the process of Rabi oscillation.

(2) For $\theta = \frac{\pi}{4}$, which means the atoms are initially prepared in the maximal entanglement state

$$|\psi(0)\rangle = \frac{\sqrt{2}}{2}(|eg\rangle + |ge\rangle), \tag{16}$$

the concurrence for atom-atom entanglement at time $t$ reduces to

$$C(\rho^{atoms}(t)) = (\frac{2r^2 + \cos\sqrt{2r^2+1}gt}{2r^2+1})^2. \tag{17}$$

In this case, the time evolution of atom-atom entanglement is relevant to the cavity-fiber coupling ($v$), the sufficient condition for concurrence (17) to be zero is $r \leq \frac{\sqrt{2}}{2}$. If $r = 0$ or $v = 0$ (there is no interaction between the cavity and fiber modes), Eq. (17) is just the same as Eq. (15).

The time evolution of entanglement of atomic system expressed in Eq. (17) is plotted in Fig.2, the results in Fig.2 (a) show that the time evolution of entanglement is sensitive to the ratio between cavity-fiber coupling ($v$) and atom-cavity coupling ($g$). This means that the entanglement dynamics can be controlled by specific $v$ and $g$.

It is also seem from Fig.2 (b) that entanglement can be strengthened by increasing the ratio $r$ which implies that if $r \to \infty (v \geq g)$ the two atoms can be maintained in the maximal entanglement state. This result can be easily explained. We have assumed that the interaction between the atoms and the cavity field and the coupling between the cavity and fiber fields are resonance. By use of the canonical transformations [12, 13]

$$a_1 = \frac{1}{2}(c_+ + c_- + \sqrt{2}c), a_2 + \frac{1}{2}(c_+ + c_- - \sqrt{2}c), b = \frac{1}{\sqrt{2}}(c_+ - c_-) \quad , \qquad (18)$$

three normal bosonic modes $c$ and $c_\pm$ are introduced. One has $c$ with frequency $\omega$, and $c_\pm$ with frequency $\omega \pm \sqrt{2}v$. The three normal modes are not coupled with each other but interact with the atoms because of the contributions of the cavity fields. However, for $r \to \infty (v \geq g)$, the interaction of atoms with the non-resonant modes is highly suppressed and the system reduces to two qubits resonantly coupled through a single harmonic oscillator.

From Fig.2 (c), we can find that all the entanglement of atoms exhibit ESD in the condition of $r \leq \frac{\sqrt{2}}{2}$, but for the case $r = \frac{\sqrt{2}}{2}$, the entanglement of atoms vanishes much slower, the effect of ESD can also be maintained for a longer time.

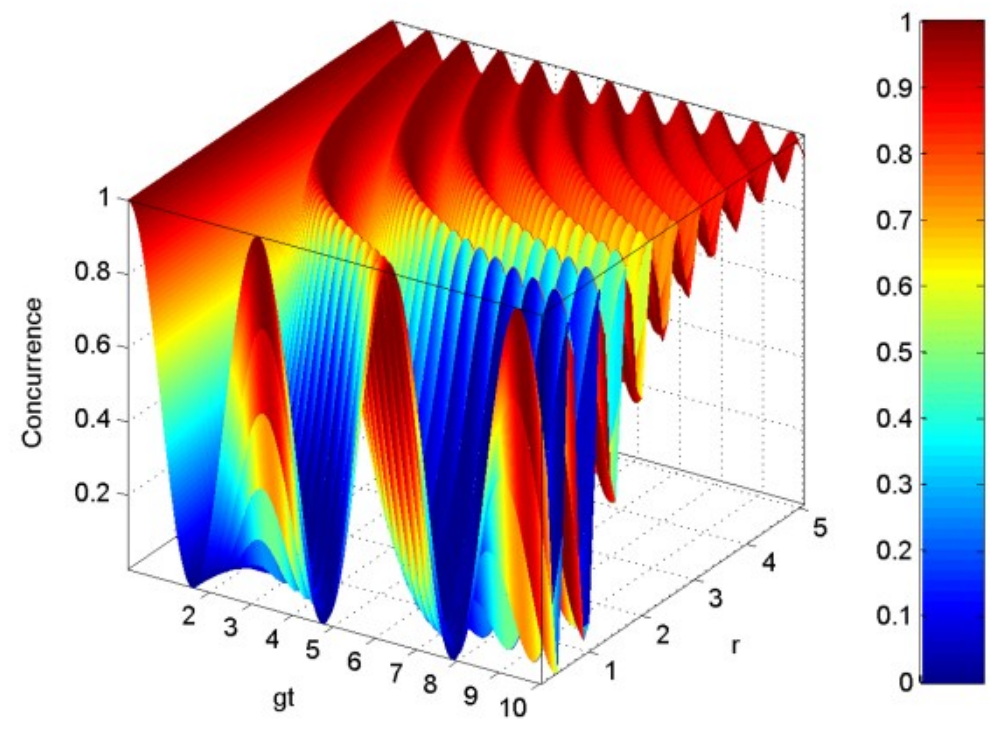

(a)

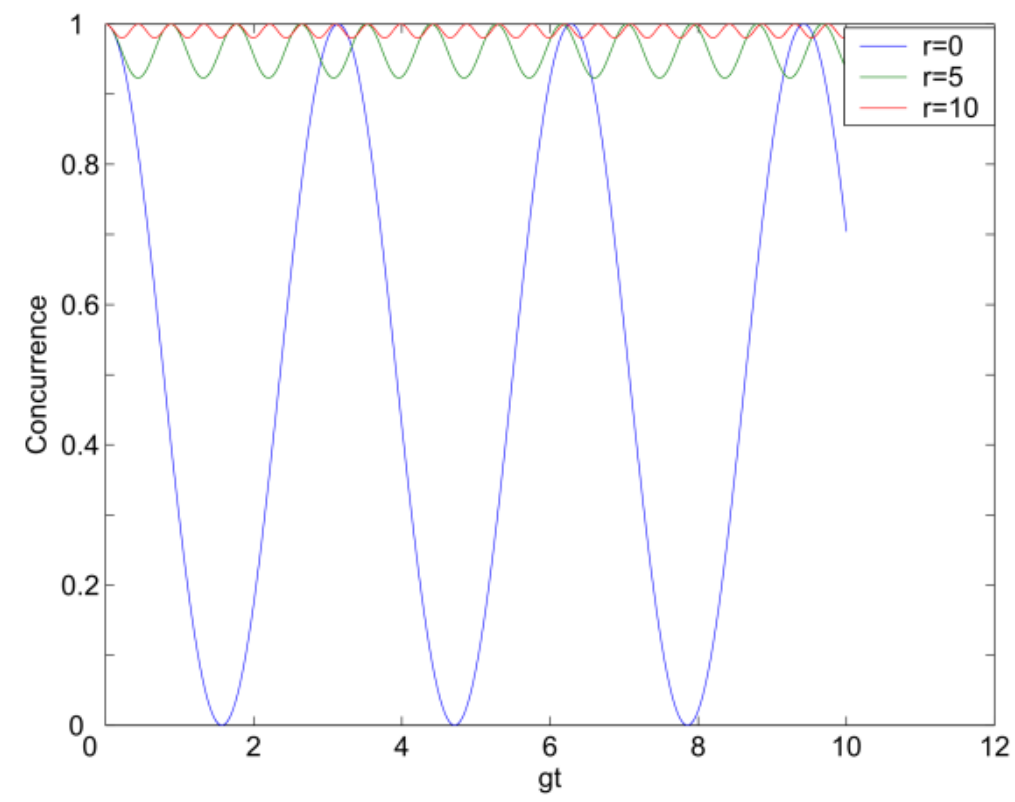

(b)

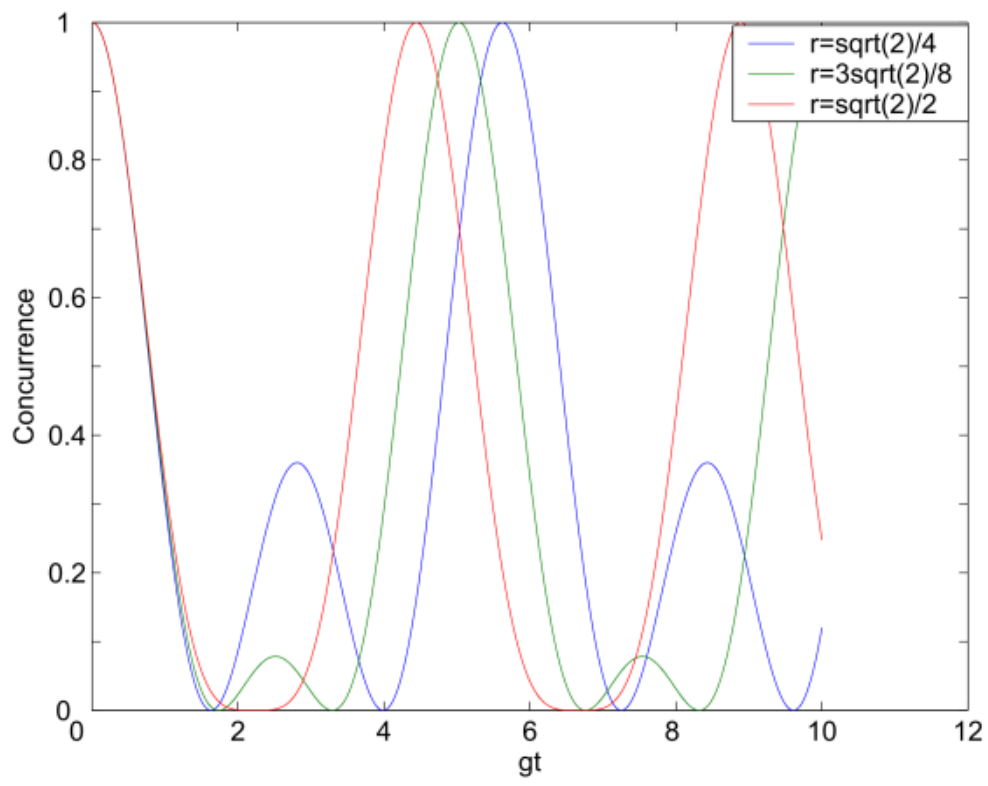

(c)

Fig.2 (Color online) the concurrence of the atoms versus the dimensionless parameters $gt$ and $r$, for the case $\theta=\frac{\pi}{4}$.

(3) For $\theta=\frac{\pi}{12}$, the atoms are initially prepared in the non-maximally entangled state. In this case, the concurrence for atom-atom entanglement at time $t$ reduces to

$$C(\rho^{atoms}(t))=2\left|[\frac{2r^2+\cos\sqrt{2r^2+1}gt}{4r^2+2}(\cos\frac{\pi}{12}+\sin\frac{\pi}{12})]^2-[\frac{\cos gt}{2}(\cos\frac{\pi}{12}-\sin\frac{\pi}{12})]^2\right|. \tag{19}$$

In Fig.3, we plot $C(\rho^{atoms}(t))$ as functions of rescaled time $gt$ and $r$ for the case $\theta=\frac{\pi}{12}$.

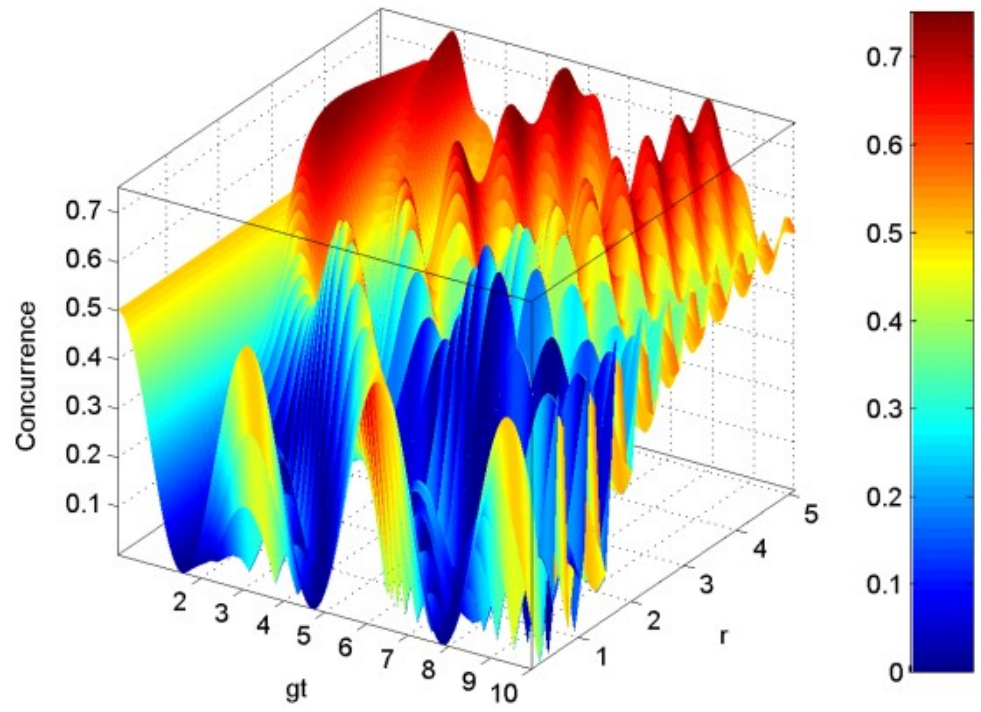

Fig.3 The concurrence of the atoms versus the dimensionless

Parameters $gt$ and $r$, for the case $\theta = \frac{\pi}{12}$.

The results from Fig.2 and Fig.3 show that the time evolution of entanglement is also sensitive to the degree of entanglement of the initial state.

Before ending this section, it is necessary to give a brief discussion on the experimental matters. The ratio $r$ between cavity-fiber coupling and atom-cavity coupling could in principle be adjusted as that of reference [26]. However, the atom-cavity coupling may be easier to modulate for it's entirely determined by the size of the Bohr orbit and the volume of the cavity mode and varies along the atom trajectory according to the law $g(z) = g_0 \exp(-z^2/w^2)$, where $z$ is the position of the atom along the beam axis [27]. It is also feasible to change the atom-cavity coupling with the assistant of classical fields similar to references [11, 19, 22, 24].

Consider a three-level atom which has one excited state $|r\rangle$ and two ground states $|e\rangle$ and $|g\rangle$. The transition $|e\rangle \rightarrow |r\rangle$ is coupled to a single cavity mode with coupling strength $g_1$ and detuning $\delta$. Meanwhile, the transition $|g\rangle \rightarrow |r\rangle$ is driven by a classical field with the Rabi frequency $\Omega$ and detuning $\delta$. Set $\delta$ □ , then the upper-lever $|r\rangle$ can be adiabatically eliminated and the atom undergoes Raman transitions. Then safely neglect the terms of cavity- and laser-induced atomic level shifts, which can be compensated by using second lasers. The system finally reduces to a model that a two-level atom interacting with a single cavity with effective coupling strength $g = \frac{g_1 \Omega}{\delta}$. Thus, coupling strength $g$ can be adjusted through changing the Rabi frequency $\Omega$.

Another issue should be concerned is that large ratio $r$ may require weak atom-cavity coupling because it is not always easy to obtain large $v$. Thus the decoherence arising from photon leakage from the cavity should be concerned in actual situation although it cannot disguise the idea that the entanglement dynamics may be controlled through the coupling constants.

## Ⅳ Conclusion

In summary, we have investigated the entanglement dynamics including the so-called sudden death for atoms trapped in distant cavities coupled by an optical fiber.

The influence of the ratio between cavity-fiber coupling ($v$) and atom-cavity coupling ($g$) and the degree of entanglement of the initial state on the ESD is also studied. The results show that the time evolution of entanglement is sensitive not only to the degree of entanglement of the initial state but also to the ratio between cavity-fiber coupling ($v$) and atom-cavity coupling ($g$). This means that the entanglement evolvement can be controlled by specific $v$ and $g$.

Acknowledgements

This work was supported by the National Natural Science Foundation of China under Grant Nos. 10974028 and Fujian Provincial Natural Science Foundation under Grant Nos. 2009J06002.